\input{aipcheck}

\documentclass[
    ,final            
  ]
  {aipproc}

\layoutstyle{6x9}

\begin{document}

\title{The Impact of Rotation on the
Evolution of Low--Mass Stars}


\author{Daniel Brown}{
  address={Astrophysics Research Institute, JMU Liverpool,
           Twelve Quays House, Egerton Wharf, Birkenhead CH41 1LD, UK,}
}

\author{Maurizio Salaris}{}

\begin{abstract}
High precision photometry and spectroscopy of low--mass stars
reveal a variety of properties standard stellar evolution 
cannot predict. Rotation, an essential ingredient of stellar evolution,
is a step towards resolving the discrepancy between model predictions
and observations.

The first rotating stellar model, continuously tracing a
low--mass star from the pre--main sequence onto the horizontal branch,
is presented. The predicted luminosity 
functions of globular clusters and surface rotation velocities
on the horizontal branch are discussed.
\end{abstract}

\maketitle


\section{Introduction}

We include rotation into the stellar evolution code FRANEC
\citep{Pietrinferni_2004}.
The stellar structure equations were corrected for the effects of
rotation according to the prescription given in \citet{Kippenhahn_1970},
allowing a 1D treatment of the stellar structure.

Additional to chemical element transport via convection
and atomic diffusion, rotational mixing (described only by the
shear instability) is introduced.
The chemical element and angular momentum transport 
induced by rotational mixing is treated as
a diffusive process with a diffusion coefficient given by
\citet{Denissenkov_2004}. The change in the total angular momentum
is accounted for during the entire evolution of the stellar model
by disk braking (Pre--Main Sequence, PMS, see \citealt{Shu_1994}), 
magnetic--wind braking (Main Sequence, MS, see \citealt{Chaboyer_1995}), 
and mass loss (Red--Giant Branch, RGB).

\section{Calibration}

Realistic initial conditions have to be found and free parameters
have to be calibrated in order to predict the rotational properties
of the stellar model and their effect on the evolution.

The initial angular rotation velocity $\omega_0$ is chosen to reproduce
the mean rotational period ($P_{\rm 1\,Myr}$) of 5\,d derived from a sample of young PMS stars 
(\citealt{Rebull_2001}, \citealt{Herbst_2002}, and
\citealt{Littlefair_2005}) in
the mass range used for calibrating surface--rotation rates for the
Hyades. To determine the initial rotation rates $\omega_0$ for
lower metallicities, the total angular momentum of stars of the
same mass, but different metallicities, is assumed to be the same 
at the start of their life as a PMS star.
Disk braking ceases at a time $\tau_{\rm DB}=5$\,Myr
after which no significant amount of circum--stellar disks can be observed.

The free parameters linked to the mixing length and the magnetic--wind braking
are calibrated comparing models with
observed properties of the Sun (e.g.~radius, luminosity, and surface
rotation rate) and
surface rotation
rates observed for the Hyades (see left panel of Fig. 1).
\begin{figure}
  \includegraphics[height=.3\textheight]{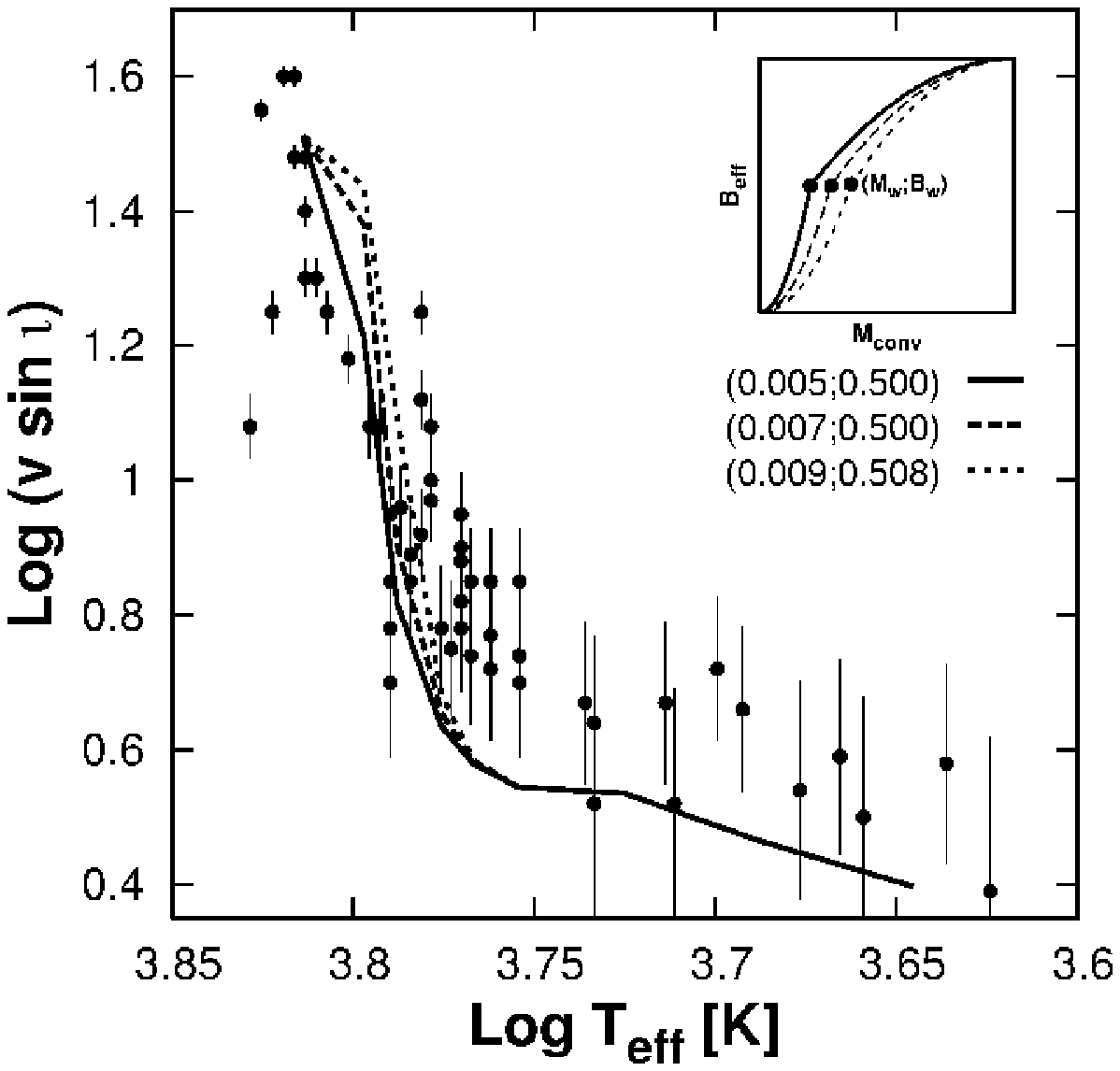}
  \includegraphics[height=.3\textheight]{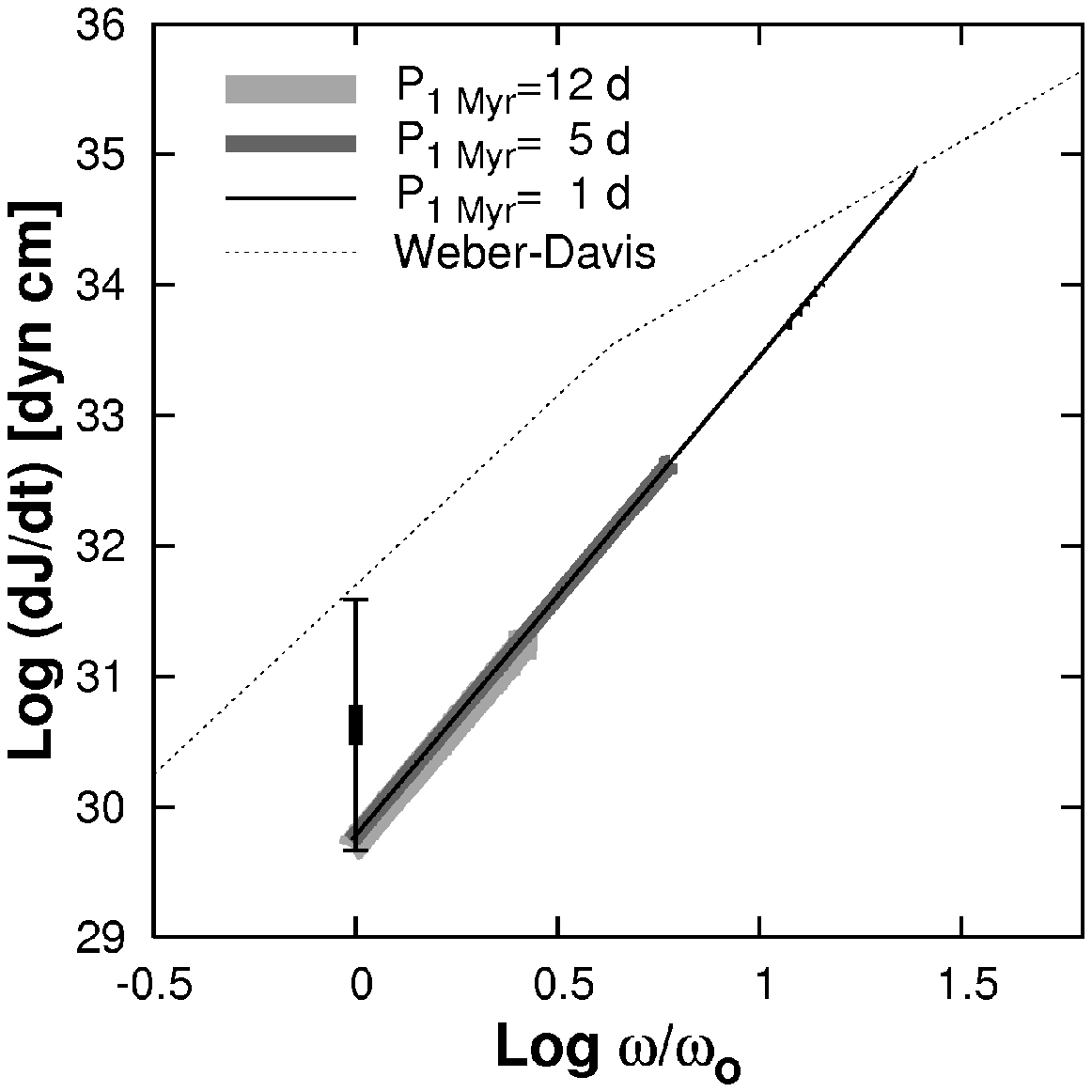}
  \caption{{\it Left:} The observed $\langle v \sin \iota \rangle$
            in the Hyades from \citet{Soderblom_1993b} as a function of 
            $T_{\rm eff}$
	    including a $\sim$1.3\,km\,s$^{-1}$
            error \citep{Soderblom_1993b} as a function of 
            $\log (T_{\rm eff})$, derived from their (B--V) colours with the
            transformation relationship by
            \citet{Alonso_1996}.
            The profile of the $B_{\rm eff}$ functions 
            describing the generation efficiency of magnetic fields
            in the magnetic--wind braking scenario are given
            in the inset, 
            including the point ($M_{\rm w}$;$B_{\rm w}$)
            that defines the shape of the $B_{\rm eff}$ function. Isochrones
            for $\langle v \sin \iota \rangle$ are shown using different
            $B_{\rm eff}$ functions:
            (0.005;0.500) solid, 
            (0.007;0.500) long dashed, and (0.009;0.508) 
            short--dashed line.
            The isochrone with (0.005;0.500) has
            the best $\chi^2$ minimisation fit to
            the data and this $B_{\rm eff}$ function
            is chosen for the further work.
	    {\it Right:} The rate of 
            angular momentum loss
            ($\frac{{\rm d}J}{{\rm d}t}$) by
            magnetic--wind braking is plotted against the
            angular surface rotation rate ($\omega$). 
            The rotational properties of the solar models with
            $\tau_{\rm DB}$=30\,Myr, but three different
            initial
            angular velocities are shown:
            $\omega_0=4.064\times 10^{-7}$\,s$^{-1}$
            ($P_{\rm 1\,Myr}$=12\,d -- light grey line),
            $\omega_0=9.740\times 10^{-7}$\,s$^{-1}$
            ($P_{\rm 1\,Myr}$=5\,d -- grey line), and
            $\omega_0=4.870\times 10^{-6}$\,s$^{-1}$
            ($P_{\rm 1\,Myr}$=1\,d -- black line).
            Observed 
            $\frac{{\rm d}J}{{\rm d}t}$ values for the Sun  
            ($\log \frac{\omega}{\omega_0}=0$) and a
            theoretically derived upper limit of a realistic 
	    2D Weber--Davis solar--wind model 
	    \citep{Weber_1967} for the MS (dotted) is shown.}
\end{figure}
The influence of changes to the magnetic--wind braking prescription 
(e.g.~the $B_{\rm eff}$ function, see inset on the left panel of Fig. 1) 
is also demonstrated by the
solid, long--dashed, and short--dashed lines in the left panel of Fig. 1.
The calibration of the braking processes is furthermore confirmed by 
comparing the 
rate of angular momentum loss of our models to the present Sun and
an upper limit of a realistic 2D Weber--Davis solar--wind model 
\citep{Weber_1967} (see right panel of Fig. 2).

\section{Results}
We aim to produce isochrones and Luminosity Functions (LFs) of globular
clusters in the Galaxy in order to derive the impact of rotation
with our realistic rotating stellar model to improve on prior exploratory
work by \citet{VandenBerg_1998}. In addition, we predict the surface
rotation velocity of Zero Age Horizontal Branch (ZAHB) stars. The recent 
increase in observed ZAHB rotation rates and our new
rotating stellar model allow improvements to the theoretical work by
\citet{Sills_2000}.
All models have a chemical composition of Z=$10^{-3}$ and Y=0.246, typical
for globular clusters in the Galaxy \citep{Zinn_1984}.

The Colour--Magnitude Diagram (CMD) of ZAHB 
is determined for varying $\omega_0$ (non--rotating,
$0.5   \times 10^{-6}$\,s$^{-1}$, and
$1.514 \times 10^{-6}$\,s$^{-1}$), given in the left panel in Fig. 2.
Rotation increases the ZAHB brightness, since faster rotating RGB stars
develop a more massive helium core at the time of the helium flash.
\begin{figure}
  \includegraphics[height=.3\textheight]{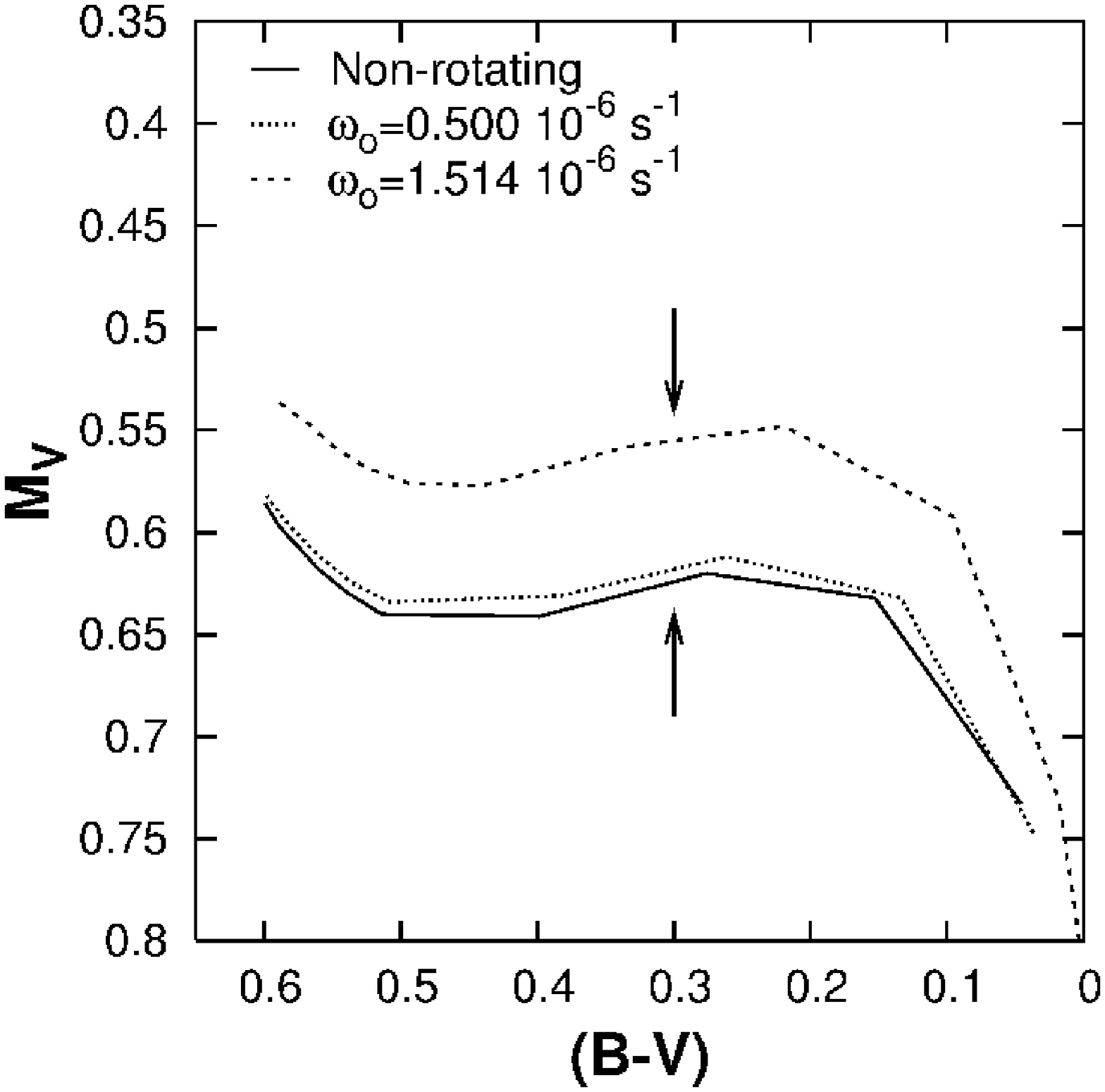}
  \includegraphics[height=.3\textheight]{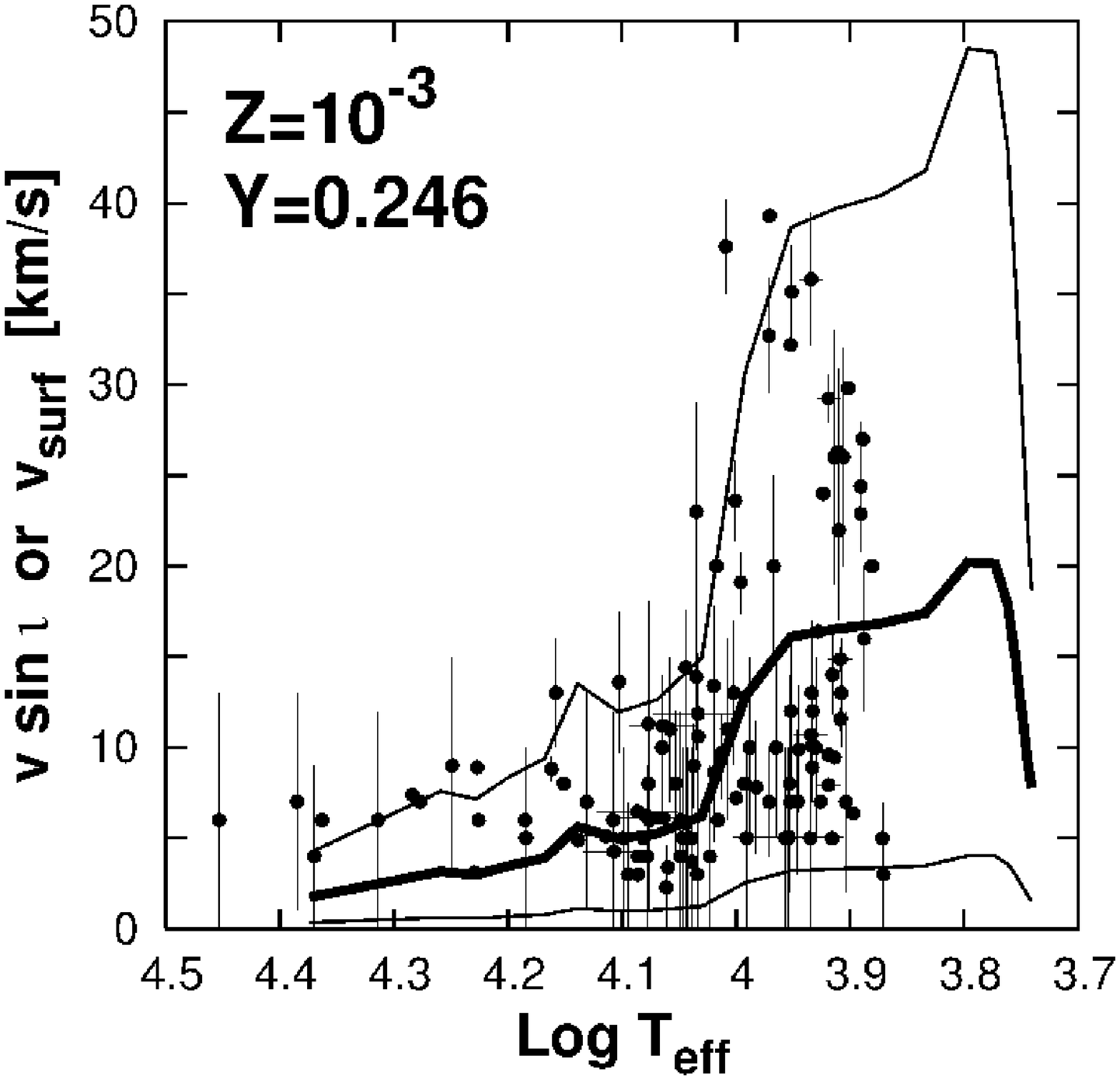}
  \caption{{\it Left:} The influence of rotation on the CMD of the 
            ZAHB (Z=$10^{-3}$, Y=0.246) is presented for three
            predicted cases of non--rotating models (solid line),
            $\omega_0=0.5   \times 10^{-6}$\,s$^{-1}$ (dotted line), and
            $\omega_0=1.514 \times 10^{-6}$\,s$^{-1}$ (dashed line).
            The ZAHB including the effects of rotation 
            ($\omega_0=1.514 \times 10^{-6}$\,s$^{-1}$) has an increased
            brightness at (B-V)=0.3\,mag (indicated by arrows) of
            $\Delta M_V=$0.07\,mag compared to the 
            non--rotating case.
           {\it Right:} The predicted ZAHB surface rotation velocities 
            given as a function of effective temperature for
            a chemical composition on the ZAHB of Z=$10^{-3}$,
            Y=0.246.
            The thick line
            represents values derived with the previously determined 
            $\omega_0$. The thin lines mark 
            the upper and lower limit of predicted ZAHB rotation values
            for initially faster or slower rotating stars, respectively. 
            Observed HB data 
            in M\,15 (\citealt{Behr_2000b} and \citealt{Recio_2004}), 
	    M\,79 \citep{Recio_2004},
            M\,13 (\citealt{Peterson_1995} and \citealt{Behr_2000a}), 
	    NGC\,2808 \citep{Recio_2004}, 
            and M\,80 \citep{Recio_2004}
            is also included.}
\end{figure}

ZAHB rotation rates are predicted using realistic initial conditions.
A relationship on the ZAHB between the surface angular velocity, 
properties of the surface convective region, and the amount of
angular momentum in the envelope around the helium core at
the tip of the RGB is determined.
It is important to note that the angular momentum within only the
surface convective region determines the surface rotation rate of
the ZAHB star.
The relationship has been derived with the predicted ZAHB rotation
rate of a 0.8\,M$_\odot$ model (without mass loss) that
has been followed through the helium flash and then extended to hotter ZAHB stars.
The predicted ZAHB rotation rates for a realistic range of $\omega_0$ 
determined from the above mentioned sample of young PMS stars
are compared with observations and given in the right panel of Fig. 2.

The impact of rotation on LFs was analysed
with four stellar models
(0.6, 0.7, 0.8, and 0.85\,M$_\odot$) from which two
LFs were generated (10.5\,Gyr
and 12\,Gyr). The young LF demonstrates that rotation
increases the brightness of the RGB bump
by 0.021\,mag
(see left panel in Fig. 3). This increase is within the scatter of
the brightness difference $\Delta_{HB}^{bump}$ between RGB bump and
HB for 
this metallicity \citep{Riello_2003}. The old LF illustrates how
rotation reduces the Sub Giant Branch (SGB) slope and increases
the number of stars at the base of the RGB by 11\,\% (right in Fig. 3).
The increase in the number count at the base of the 
RGB is still detectable (5\,\%)
when comparing the non--rotating 12\,Gyr old and rotating 10.5\,Gyr old LF,
where both cases are in agreement with observational constraints.
\begin{figure}
  \includegraphics[height=.3\textheight]{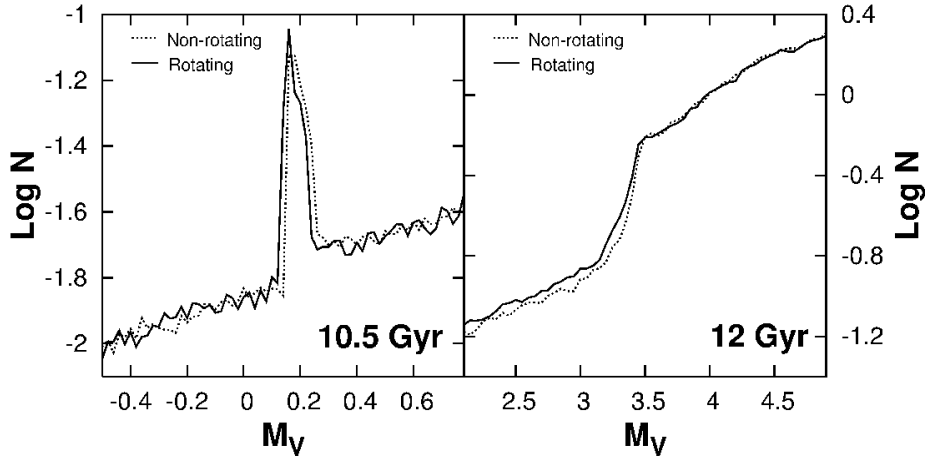}
  \caption{LFs for initial chemical composition
            of Z=$10^{-3}$, Y=0.246 are shown.
            {\it Left:}
            Two LFs of the RGB bump are shown for a 10.5\,Gyr 
            old isochrone. The first includes
            rotation ($\omega_0=1.514 \times 10^{-6}$\,s$^{-1}$; solid line) 
            and the second does not (dotted line). Rotation increases
            the brightness of the RGB bump by
            $\Delta M_V=$0.021\,mag.
            {\it Right:}
            12\,Gyr old LFs are shown. The LF including rotation is
            plotted as solid lines and the non--rotating
            comparison is plotted as a dotted line.
            The SGB shows a less steep change
            in the LF when rotation is included, leading
            to a 11\,\% increase in number of stars at the base of the RGB.}
\end{figure}

\section{Summary}
The rotating models demonstrate the brightening of the ZAHB and RGB
bump due to rotation. A $\omega_0=1.514 \times 10^{-6}$\,s$^{-1}$
results in an increased brightness of the ZAHB of $\Delta M_V=0.07$\,mag 
and an increased brightness of the RGB bump of $\Delta M_V=0.021$\,mag
compared to the non-rotating case.
The increase in the helium--core mass
and number count at the base of the RGB are just two effects of
stellar rotation on low--mass stars that will be addressed in
future work.

The rotating code and its capability to follow the 
evolution through the helium flash makes it the first
stellar evolution code to include rotation during the helium--flash
phase for low--mass stars.
Given realistic initial rotation rates, the code
allows us to predict the ZAHB
velocity distribution.
These predicted rotation rates agree better
with observations than rates determined in
the most recent theoretical work by \citet{Sills_2000}.
Our determined ZAHB surface rotation rates explain the
apparent ``bimodality'' of the rotation rates at 11\,000\,K (see right panel in
Fig. 2) without the need of a bimodality in the initial rotation rates
and its survival up to the tip of the RGB.



\bibliographystyle{mn2e}

\bibliography{proceeding_cam}

\end{document}